\documentclass{article}

\usepackage{PRIMEarxiv}

\usepackage[utf8]{inputenc} 
\usepackage[T1]{fontenc}    
\usepackage{hyperref}       
\usepackage{url}            
\usepackage{booktabs}       
\usepackage{amsfonts}       
\usepackage{nicefrac}       
\usepackage{microtype}      
\usepackage{lipsum}
\usepackage{fancyhdr}       
\usepackage{graphicx}       
\graphicspath{{media/}}     

\usepackage{bm}

\title{X-ray Dissectography Enables Stereotography to Improve Diagnostic Performance
}

\author{
  Chuang Niu and Ge Wang \\
  AI-based X-ray Imaging System (AXIS) Lab, Department of Biomedical Engineering\\
  Rensselaer Polytechnic Institute, 110 8th Street, Troy, New York 12180, USA \\
  \texttt{\{niuc, wangg6\}@rpi.edu} \\
}

\begin{document}
\maketitle

\begin{abstract}

X-ray imaging is the most popular medical imaging technology. While x-ray radiography is rather cost-effective, tissue structures are superimposed along the x-ray paths. On the other hand, computed tomography (CT) reconstructs internal structures but CT increases radiation dose, is complicated and expensive. Here we propose "x-ray dissectography" to extract a target organ/tissue digitally from few radiographic projections for stereographic and tomographic analysis in the deep learning framework. As an exemplary embodiment, we propose a general X-ray dissectography network, a dedicated X-ray stereotography network, and the X-ray imaging systems to implement these functionalities. Our experiments show that x-ray stereography can be achieved of an isolated organ such as the lungs in this case, suggesting the feasibility of transforming conventional radiographic reading to the stereographic examination of the isolated organ, which potentially allows higher sensitivity and specificity, and even tomographic visualization of the target. With further improvements, x-ray dissectography promises to be a new x-ray imaging modality for CT-grade diagnosis at radiation dose and system cost comparable to that of radiographic or tomosynthetic imaging.

\end{abstract}


\section{Introduction}
\label{sec_intro}

X-ray imaging is the first and still most popular modern medical imaging approach, which is performed by various kinds of systems. In the low-end, x-ray radiography takes a two-dimensional projective image through a patient, which is called a radiogram or radiograph. In the high end, many x-ray projections are first collected and then reconstructed into tomographic images transversely or volumetrically. Between these two extremes, digital tomosynthesis takes a limited number of projections over a relatively short scanning trajectory and infers three-dimensional features inside a patient. These x-ray imaging modes have their strengths and weaknesses. X-ray radiography is cost-effective but it produces a single projection, and has multiple oranges and tissues superimposed along x-ray paths, compromising the diagnostic performance. On the other hand, x-ray CT unravels structures overlapped in the projection domain into tomographic images in a 3D coordinate system but CT uses a much higher radiation dose, is complicated and expensive. Digital tomosynthesis is a balance between x-ray radiography and CT in terms of the number of needed projections, the information in resultant images, and the cost to build and operate the imaging system.

Reducing radiation dose and improving imaging quality and speed are the main tasks for the development of x-ray imaging technologies.
As x-ray radiography has the lowest radiation dose, the fastest imaging speed, and the lowest price, researchers have been focusing on improving the radiogram quality. Currently, there are mainly two ways for this purpose: 1) suppressing interfered structures \cite{stes, ribcenterline, li2019encoding, peng2020xraysyn, hbs, han2021ganbased} or enhancing related structures \cite{enhance, li2019encoding}, and 2) generating 3D volumes \cite{Ying_2019_CVPR, shen2019patient, shen2021geometryinformed}.
It is well known that superimposed anatomical organs/tissues in 2D radiographs significantly complicate signal detection, such as for diagnosis of lung diseases.
To address this challenge, in early studies \cite{stes, ribcenterline}  model-based methods were developed to suppress ribs in chest radiographs, some of which require manually annotated bone
masks. In recent years, deep learning methods \cite{li2019encoding, peng2020xraysyn, hbs, han2021ganbased} were proposed for suppression of ribs by leveraging 3D CT prior. Specifically, Zhou's group proposed DecGAN \cite{li2019encoding} to decompose Chest X-Ray (CXR) images into different components with unpaired supervision aided by CT volumes, and the outputs show suppressed interference and enhanced target tissues. Recently, they further extended DecGAN for processing high-resolution CXR images \cite{hbs}. Along this line, they proposed a Rib Suppression GAN (RSGAN) model using a disentanglement learning scheme \cite{han2021ganbased}.
Peng et al. designed an XraySyn model that first generates a 3D volume from a single 2D x-ray image and then maps the generated volume to 2D new radiographs with bones suppressed.
Instead of suppressing the ribs in CXR images, Gozes and Greenspan proposed to enhance lung structures by extracting the extracted lungs first and then adding the result back with a scaling factor \cite{enhance}.
Generating a 3D volume from a single or a few radiographs is another way to improve radiography.
Ying et al. proposed X2CT that generates a 3D CT volume from a pair of orthogonal radiographs using a CycleGAN framework \cite{Ying_2019_CVPR}.
Recently, Xing's group proposed to generate a 3D volume from single or few radiograhps.
Specifically, their GAN-based model is for patient-specific volume generation from a single projection \cite{shen2019patient}.
Also, they proposed a geometry-informed GAN-based framework to generate a 3D volume from ultra-sparse views \cite{shen2021geometryinformed}.
This method works in the three steps: 1) a disentanglement GAN-based model to synthesize new views; 2) Backprojection of all projections to form a 3D volume; 3) A 3D GAN network to refine the volume.

Among the above-surveyed methods, those methods that are for suppressing/enhancing specific structures are mainly intended to improve the performance of classification \cite{li2019encoding, peng2020xraysyn, hbs, han2021ganbased} and detection \cite{enhance} from a single radiograph without providing 3D information. On the other hand, although the rest of the existing methods that map 2D radiograms to 3D CT volumes achieved remarkable results, they cannot reconstruct structures accurately and reliably, since their clinical utilities have not been demonstrated so far.
Particularly, it can be seen that almost all the above methods depend on the GAN framework and/or unpaired learning for 2D/3D image generation. A major potential problem is that GAN-based models tend to generate fake structures, which is a major concern in the medical imaging field.

In this study, we propose x-ray dissectography (XDT) in general and specialize it for x-ray stereotography to improve image quality and diagnostic performance. The essential idea is that we digitally extract a target organ/tissue from the original radiograph or radiogram which contains superimposed organs/tissues via deep learning, facilitating both visual inspection and quantitative analysis.
Considering that radiograhs from different views contain complementary information, we design a physics-based XDT network to extract the multi-view features and transform them into a 3D space. In this way, the target organ can be synergistically analyzed in isolation from different projection angles.
As a special yet important application of our XDT approach, we propose X-ray stereography that allows a reader immersively perceive the target organ/tissue from two dissected radiographs in 3D, synergizing machine intelligence and human intelligence, similar to what CT does.
Biologically, stereo perception is based on binocular vision for the brain to reconstruct a 3D scene, and can be applied to see through dissected radiograms and form a 3D rendering in a radiologist in mind.
In this work, we design an X-ray imaging system dedicated to this scenario. Different from our daily visual information processing, which senses surroundings with reflected light signals, radiograms are projective through an object to allow 3D concept of x-ray semi-transparent features.

To avoid fake structures, we optimize XDT neural networks in a supervised 2D-to-2D learning paradigm without using a GAN-like model.
However, it is not feasible to obtain ground-truth radiographs of a separated organ of interest in a living patient.
Hence, we use  widely available CT volumes reconstructed from a sufficiently large number of radiograms in different projection angles in the training stage.
To obtain a 2D radiograph of a target organ without surrounding tissues, we can manually or automatically segment the organ in the associated CT volume first and then compute the ground-truth radiograph through projecting the dissected organ according to the system parameters.
In other words, radiographs and CT images are obtained from the same patient and the same imaging system to avoid unpaired learning.
In practice, paired 2D radiographs and CT volumes can be easily obtained on any CT system since cone-beam projections are essentially 2D radiographs. Also, we can collect such data or images using other systems including the twin robotic x-ray system \cite{robotic}.
On the other hand, over the past years numerical simulation tools \cite{dukesim, catsim} and digital phantoms \cite{xcat} have become mature and play an important role in a wide range of applications.
In this context, we can utilize a cutting-edge simulation platform, such as the popular academic \cite{dukesim} and industrial \cite{catsim} simulators, for training XDT networks. These simulators can take either a clinical CT volume or a digital 3D phantom to compute a conventional x-ray radiograph, and then extracts a target organ/tissue digitally to produce the ground-truth radiograph of the organ/tissue.
In real applications, we can use domain adaption techniques to optimize the performance of our XDT networks by integrating both simulated and real datasets.
Our initial experimental results have shown that XDT indeed separates the lungs with faithful texture and structures, and we can perceive the extracted lungs via stereoscopic viewing with a pair of 3D glasses. Potentially, this approach can improve the diagnostic performance in lung cancer screening, COVID-19 follow-up, and other applications.

The rest of this paper is organized as follows. In the second section, we describe the deep networks and related key aspects. In the third section, we report our experimental results. In the last section, we discuss several issues and conclude the paper.

\section{Methodology}
\subsection{General Workflow of X-ray Dissectography}
\begin{figure}
    \centering
    \includegraphics[width=1.0\textwidth]{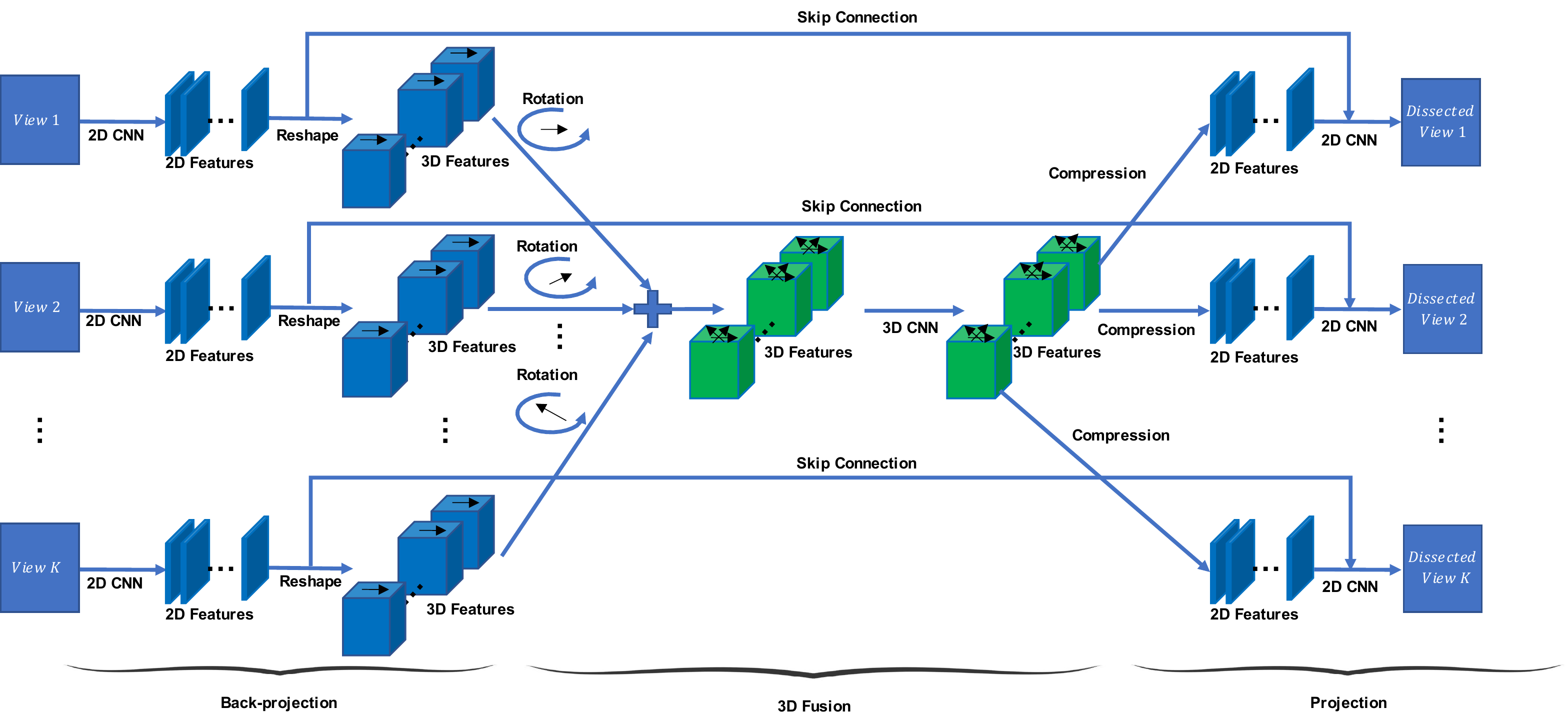}
    \caption{X-ray dissectography network (XDT-Net).}
    \label{fig_xdt}
\end{figure}

X-ray dissectography (XDT) is dedicated to transform a conventional radiogram $\bm{x} = \sum_{i=1}^B \bm{y}_i + \bm{y}_t$ to a projection image $\bm{y}_t$, where $\bm{y}_t$ is a projection of only a target organ/tissue, and $\sum_{i=1}^B \bm{y}_i$ represents the superimposed image of the $B$ anatomical components involved in the conventional radiogram. In fact, it is an extremely ill-posed problem as we can only observe the radiograph $\bm{x}$, and impossible to obtain any analytic solution in a general setting.
Fortunately, a specific organ/tissue in the human body has a fixed relative location, a strong prior on material composition, and similar patterns (such as shapes, textures, and other properties).
Given this kind of knowledge, a radiologist can identify different organs/tissues in the conventional radiogram.
However, the superimposed organs/tissues challenge the human in visual inspection for the target one.
Here we leverage deep neural networks (DNNs) to learn such priors and extract purified radiographs as if x-rays go only through the target organ.
Such DNNs can be trained with an individual or a specific population for quantitative accuracy and clinical utilities.

In this study, we propose a physics-based XDT network (XDT-Net) for separating a target organ/tissue from more than one views, as shown in Fig. \ref{fig_xdt}. 
Note that various organs/tissues may be separated using this framework in different combinations, depending on specific applications.
The XDT-Net consists of the three modules: 1) a back-projection module, 2) a 3D fusion module, and 3) a projection module.
The back-projection module maps 2D radiographs to 3D features, like a tomographic back-projection process.
It consists of $k$ 2D convolutional neural networks (CNNs) followed by reshape operators, where $k$ is the number of input views, each CNN is applied to a specific view, and different CNNs have the same architecture but trainable parameters may be optimized differently.
The fusion module integrates the information from all views in the 3D feature space. It first aligns the 3D features of different views by rotation according to their projection angles, and then combines them by a 3D CNN.
The projection module predicts each radiograph containing only the target organ/tissue. It first squeezes each 3D feature volume to a 2D feature map along a given angle, and then the 2D CNN takes the 2D features from both the squeeze operator and the back-projection module to predict the radiograph of the target organ only.

\subsection{Specific Embodiment for X-ray Stereography}
\label{sec_xst}

\begin{figure}
    \centering
    \includegraphics[width=1.0\textwidth]{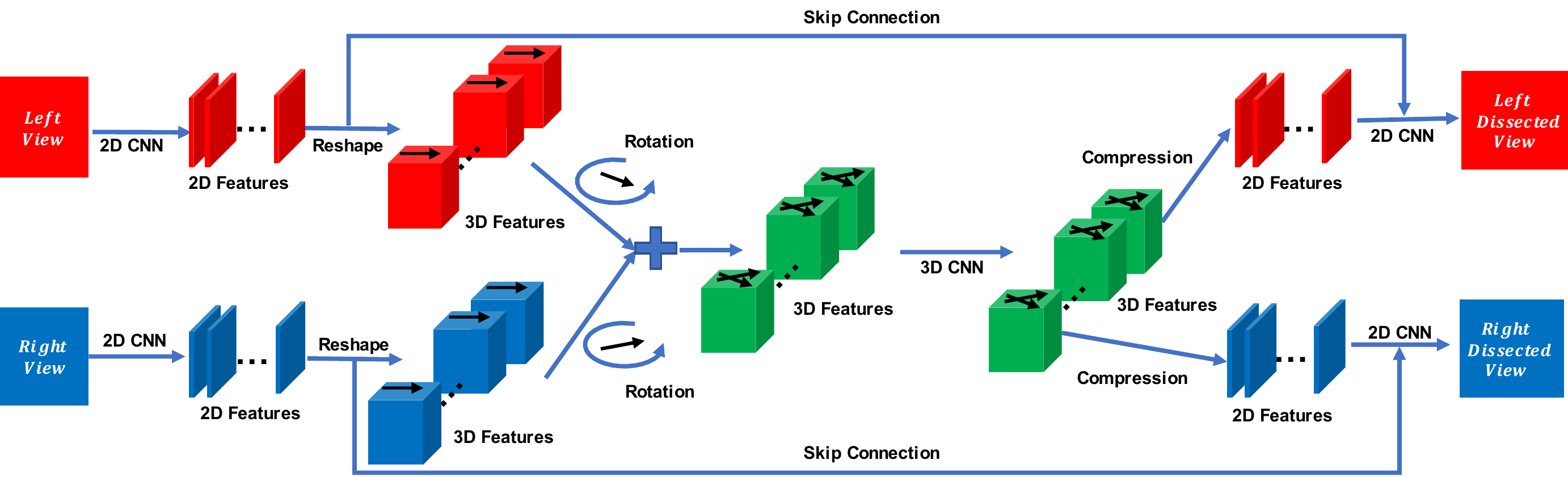}
    \caption{X-ray stereoraphic imaging network (XST-Net).}
    \label{fig_xst}
\end{figure}

We perceive the world in 3D thanks to binocular vision. Given binocular disparity, the human brain is capable of sensing the depth in the scene. Inspired by this amazing fact, here we investigate X-ray stereography (XST) with two radiograms of an isolated organ. When inspecting the human body with x-rays, organs/tissues with large linear attenuation coefficients will overwhelm the ones with small attenuation coefficients in radiograms.
As a result, it is difficult to discern subtle changes in internal organs/tissues due to the superimposition of multiple organs/tissues, significantly compromising stereopsis. With our proposed XDT-Net and XST-Net, we can integrate  machine intelligence for target organ dissection and human intelligence for stereographic perception so that a radiologist can perceive a target organ in 3D with details much more vivid than in 2D, potentially improving diagnostic performance.

To enable XST of a specific organ, we adapt the XDT-Net to the XST-Net as shown in Fig \ref{fig_xst}.
The XST-Net also consists of the same three modules: the backprojection module, the 3D fusion module, and the projection module.
Each module of the XST-Net shares the same network architecture as that of the XDT-Net but needs to be adapted for stereo viewing.
First, the backprojection module of the XST-Net takes two radiographs as inputs, which are two images into our eyes.
Second, in reference to the view angles of two eyes, the 3D fusion module uses a different rotation center to align 3D features from two branches appropriately.
Our proposed XST imaging system is described in Subsection \ref{sec_system}.
Third, the projection module translates the merged 3D feature first and then squeezes it to 2D feature maps according to the human reader's viewing angles.
Finally, two dissected radiographs are respectively sent to the left and right eyes through a pair of 3D glasses for stereoscopy.

\subsection{Design Considerations on X-ray Dissectography and Stereography}
\label{sec_system}

\begin{figure}
    \centering
    \includegraphics[width=1.0\textwidth]{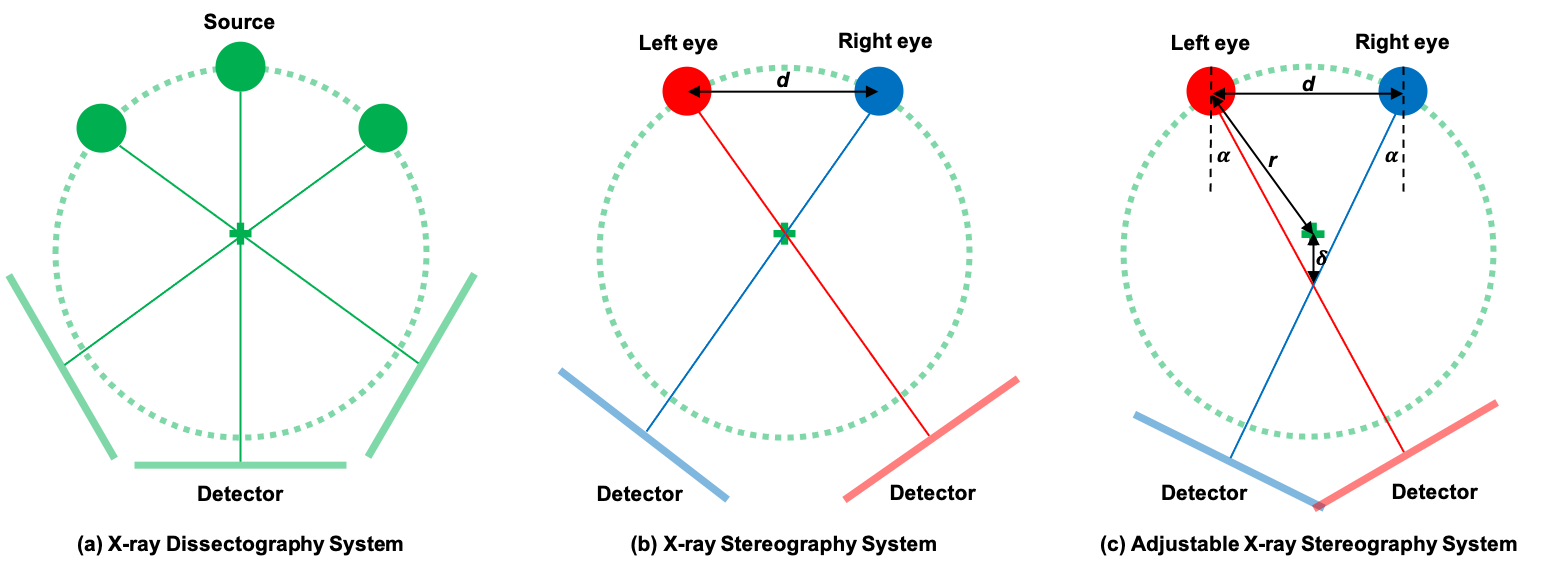}
    \caption{X-ray imaging system configuration to facilitate radiographic, stereographic and tomographic analysis on a digitally isolated organ or tissue.}
    \label{fig_system}
\end{figure}

We assume that radiographs from sufficiently many different angles can be obtained in the network training stage such that image volumes can be reconstructed.
The traditional cone-beam CT system, as shown in Fig. \ref{fig_system} (a), serves this purpose. Then, many pairs of conventional radiographs and the counterparts of the target-only radiographs can be obtained from a reconstructed CT volume and a segmented organ in the reconstructed CT volume respectively. In the testing stage, the same XDT system only needs to generate few radiographs at any angles for the trained XDT-Net to extract the corresponding radiograms of the target organ/tissue alone, without surround tissues, for much-improved visual inspection and quantitative analysis.
To achieve x-ray stereopsis, we design an XST imaging system, as shown in Fig. \ref{fig_system} (b) and (c), where each source is regarded as an eye while the projection through the body is recorded on the opposite detector.
In a simple setting, we directly take two radiograms from the XDT system so that the distance between eyes is $d$, as shown in Fig. \ref{fig_system} (b). In this case, the center X-rays from the source positions intersect at the object center.
For the adaption to different applications and readers, we further design an adjustable XST system in Fig.~\ref{fig_system}~(c).
There are two parameters of the system to control the offset between the two eyes and the viewing angle from a pre-specified principal direction.
In Fig. \ref{fig_system}~(c), red and blue dots denote the left and right eyes, the red and blue plates are the corresponding detectors, and green cross is viewed as the object center.
The distance between two eyes is $d$, the distance between the source and the object center is $r$, and the angle between center X-ray and the pre-specified reference direction is $\alpha$ for both eyes. Thus, the intersection point of two center X-rays is translated from the object center along the vertical direction.
The distance offset $\delta$ can be computed as $\delta = \frac{d}{2\tan(\alpha)} - \sqrt{r^2 - (d/2)^2}$, which is used to adjust the rotation center for XST-Net as introduced in Subsection \ref{sec_xst}.

In practice, the XST system for inspecting different organs/tissues may require different geometric parameters. Both XDT and XST systems can be implemented in various ways such as with robotic arms \cite{robotic} so that the geometric parameters can be easily set to match a reader's preference.

\section{Experimental Results}

\subsection{XDT and XST Simulation}

To train the proposed XDT and XST networks, how to obtain sufficient  training data is critically important. In this study, we use the  X-ray simulation tools and digital phantoms to generate such data for training models and evaluating the feasibility and utilities of these systems and methods.
Here we used a clinical CT dataset, denoted as CT-lung, and simulated radiograms in cone-beam geometry.
Specifically, 50 reconstructed CT volume of patients were selected, including 10 patients from \cite{mosdata} and 40 patients from \cite{orgdata}.
The data from \cite{mosdata} provide the 3D lung masks. Hence, we can simulate the paired radiograms with and without lung masks \cite{astra}.
Note that before performing the cone-beam projection, the patient bed in the CT volume was first masked out in a semi-automatic manner using MITK software.
Since the lung masks provided in \cite{orgdata} are not consistent with those provided in \cite{mosdata}, we trained a semantic segmentation UNet with the annotated data in \cite{mosdata} to identify the body region for removal of the patient bed and segment the lung region slice-by-slice for the data in \cite{orgdata} consistently.
When 2D radiograms were synthesized, we rotated the patient CT volume from 0$^{\circ}$ to 180$^{\circ}$, where the angle of 0$^{\circ}$ is the frontal view.
Totally, we obtained 9,000 pairs of radiograms, and the image size is $320 \times 320$.

In this feasibility study, we evaluated not only the generalizability to different viewing angles for the same patients but also the generalizability to different patients.
In the former case, the projection angles in $[0, 6, \cdots, 174]$ for each patient were selected for training, and the angles in $[3, 9, \cdots, 177]$ were used for testing. In the latter case, we randomly selected two patients for testing, and the rest patients for training.

\subsection{XDT Results}
\label{sec_result_xdt}

\begin{figure}
    \centering
    \includegraphics[width=1.0\textwidth]{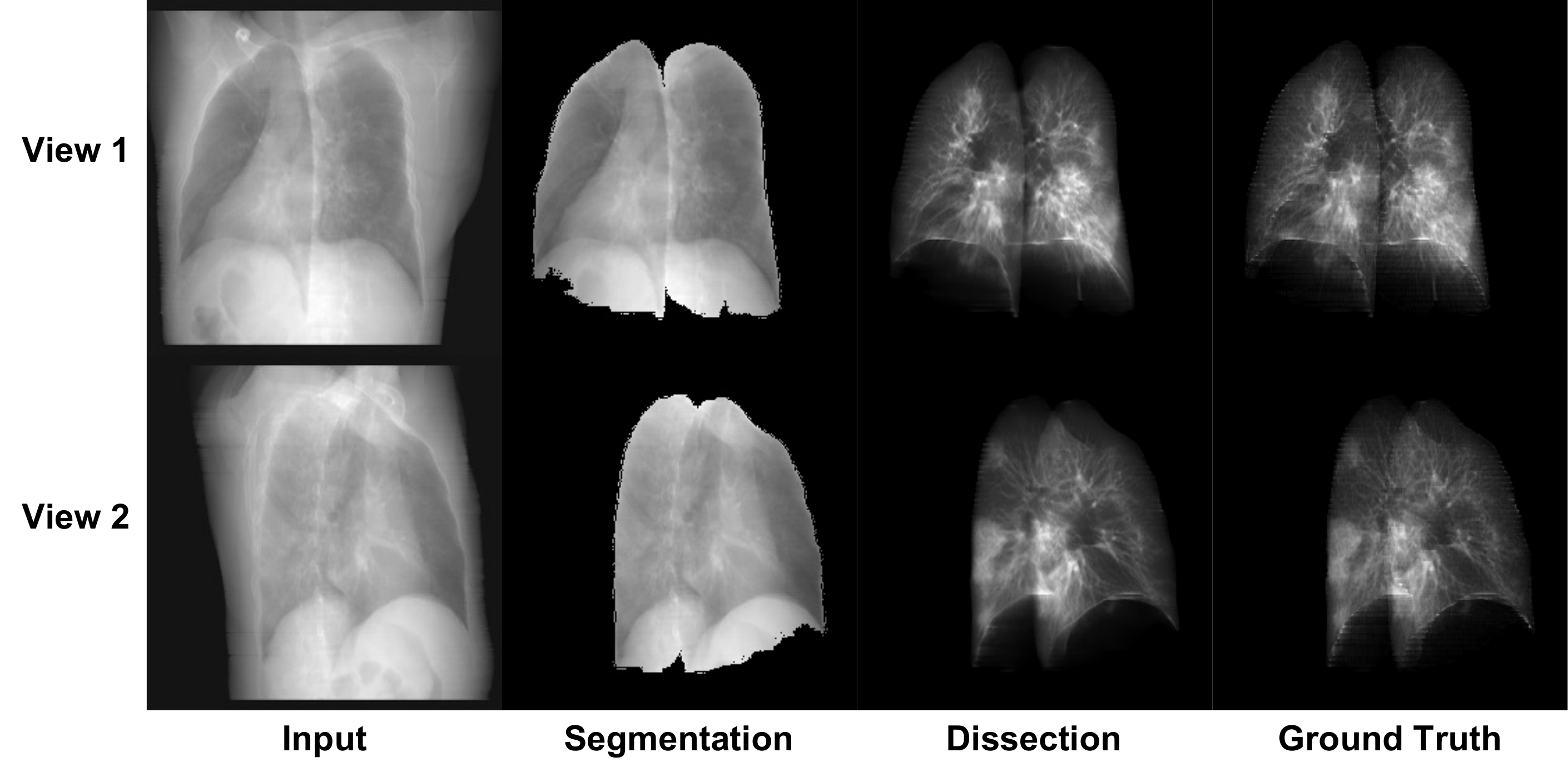}
    \caption{XDT testing results from two new orthogonal views of the same patient.}
    \label{fig_result_angle_xdt}
\end{figure}

\begin{figure}
    \centering
    \includegraphics[width=1.0\textwidth]{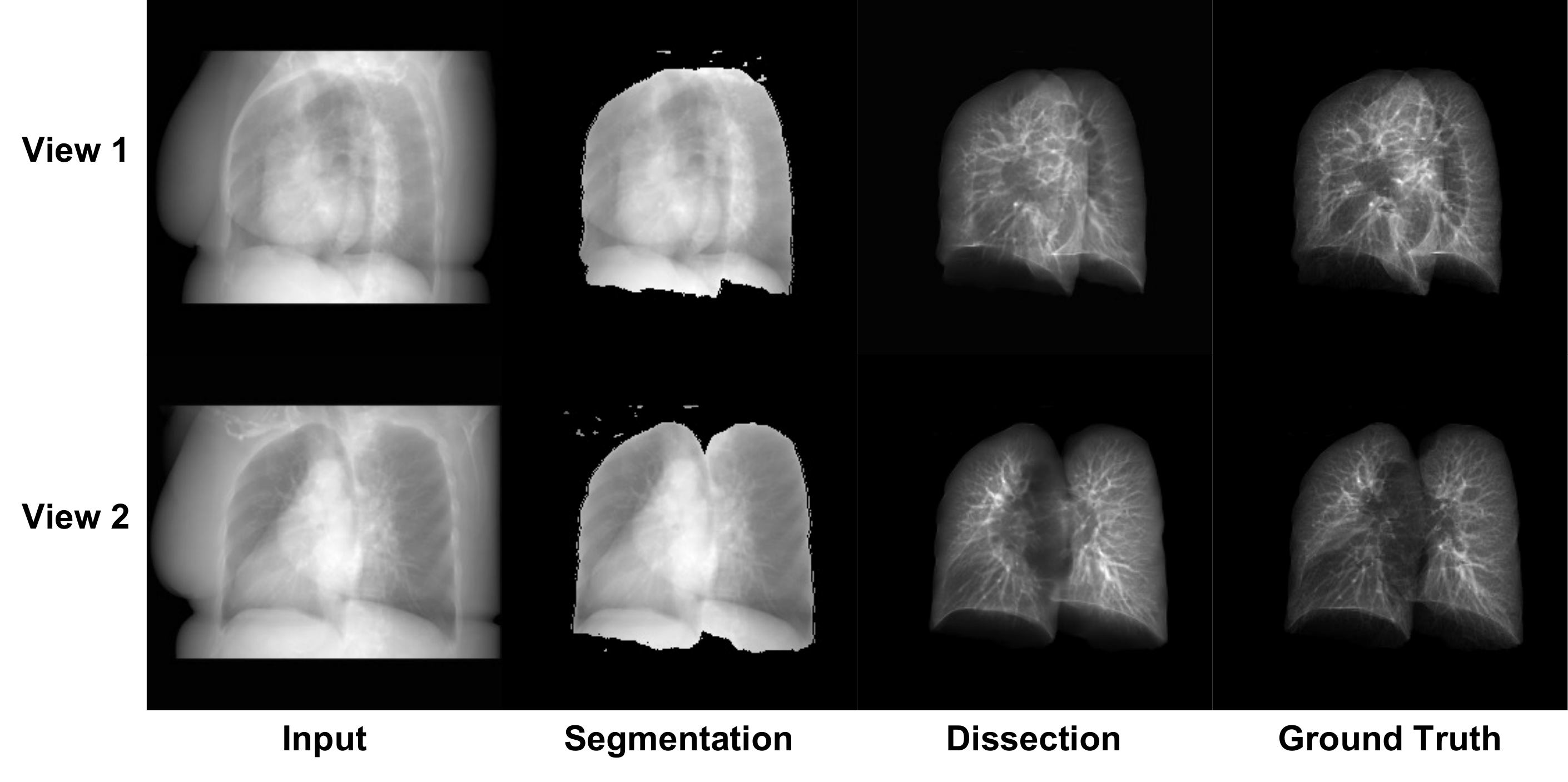}
    \caption{XDT testing results from two orthogonal views of a different patient, who was not used in the training stage.}
    \label{fig_result_pat_xdt}
\end{figure}

We first evaluated the effectiveness of the proposed XDT-Net on the CT-lung dataset.
In the current experiments, we focused on simultaneously dissecting two radiograms at orthogonal angles.
To be more specific, the region of interest was first segmented from 2D radiograms before forwarding to the XDT-Net.
In this way, the task of XDT-Net is purified to improve dissection results.
For this purpose, we trained a segmentation model to identify the region of interest.
The target mask for training this segmentation model can be easily obtained by thresholding the radiogram of isolated lungs, where the threshold was empirically set to 0.01 in the unit of linear attenuation coefficient.

The testing results are shown in Fig. \ref{fig_result_angle_xdt}, where radiographs of the same patient were collected from different angles. From the first to the fourth columns are the input radiograms, the segmentation results, the dissection results, and the ground-truth respectively. The first and the second rows present two orthogonal projections respectively.
The visual inspection shows that the dissected radiograms are very close to the ground-truth in terms of detailed structures despite being slightly smoother.
The blurring effect may be due to noise reduction  \cite{niu2021suppression}.
Compared to normal radiograms, the dissected radiograms remove irrelevant surrounding structures, highlight the target organ, and potentially improve the diagnosis performance.
Also, we evaluated the generalizability of XDT-Net to different patients.
As shown in Fig. \ref{fig_result_pat_xdt}, the results show that the dissected radiograms of new patients are not as good as those for the scanned patients (i.e., the patient has a prior CT scan).
Nevertheless, most of the extracted structures are still the same as the ground truth, although some of detailed structures are in different contrasts.
This problem is most likely due to a small (48 patients) dataset used for training in this feasibility demonstration.
We believe that the generalizability can be significantly improved if  a much larger training dataset is used; say thousands or even millions of patients. 

\subsection{XST Results}

\begin{figure}
    \centering
    \includegraphics[width=1.0\textwidth]{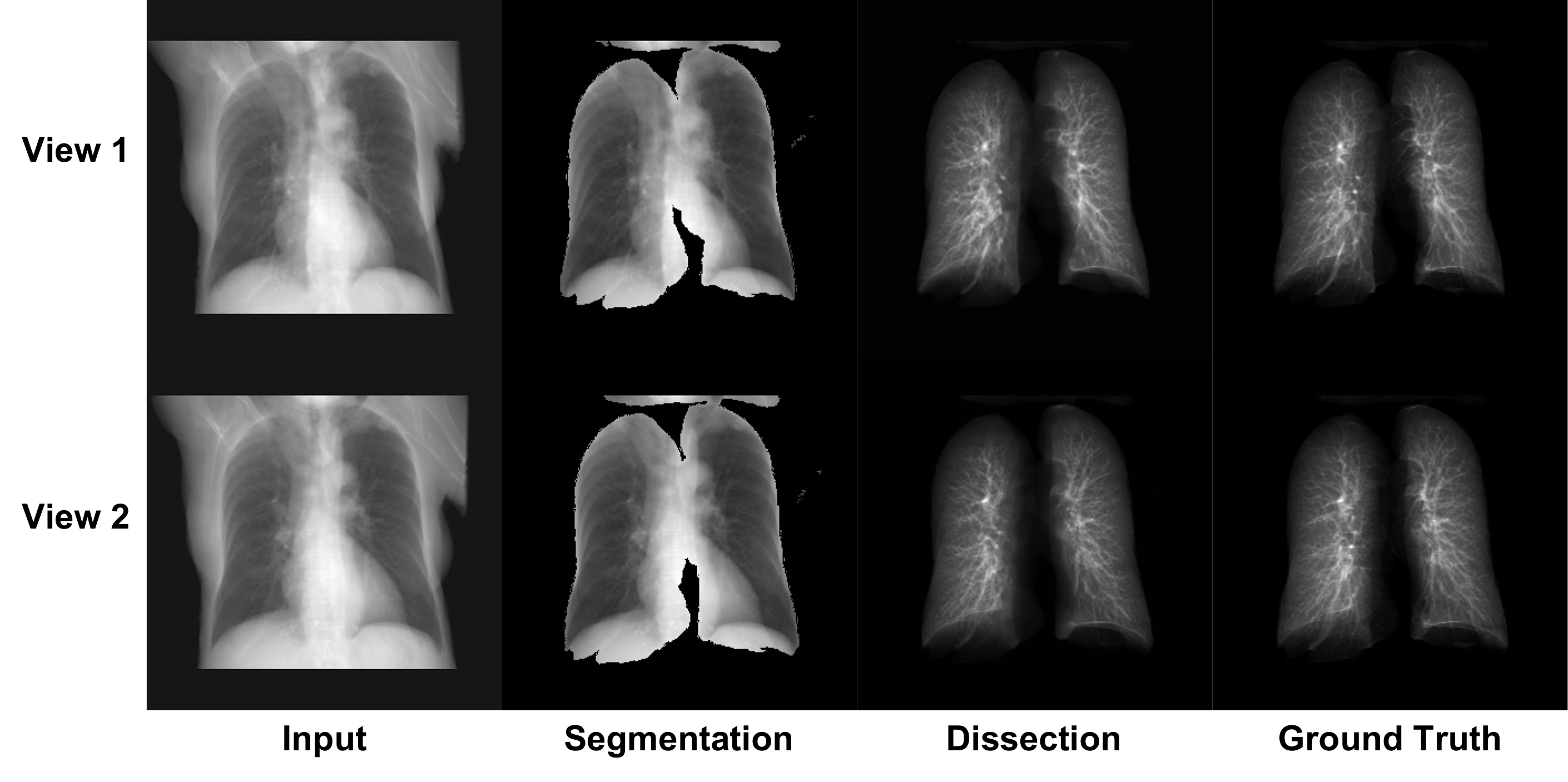}
    \caption{XST testing results from two new stereo views of the same patient.}
    \label{fig_result_angle_xst}
\end{figure}

\begin{figure}
    \centering
    \includegraphics[width=1.0\textwidth]{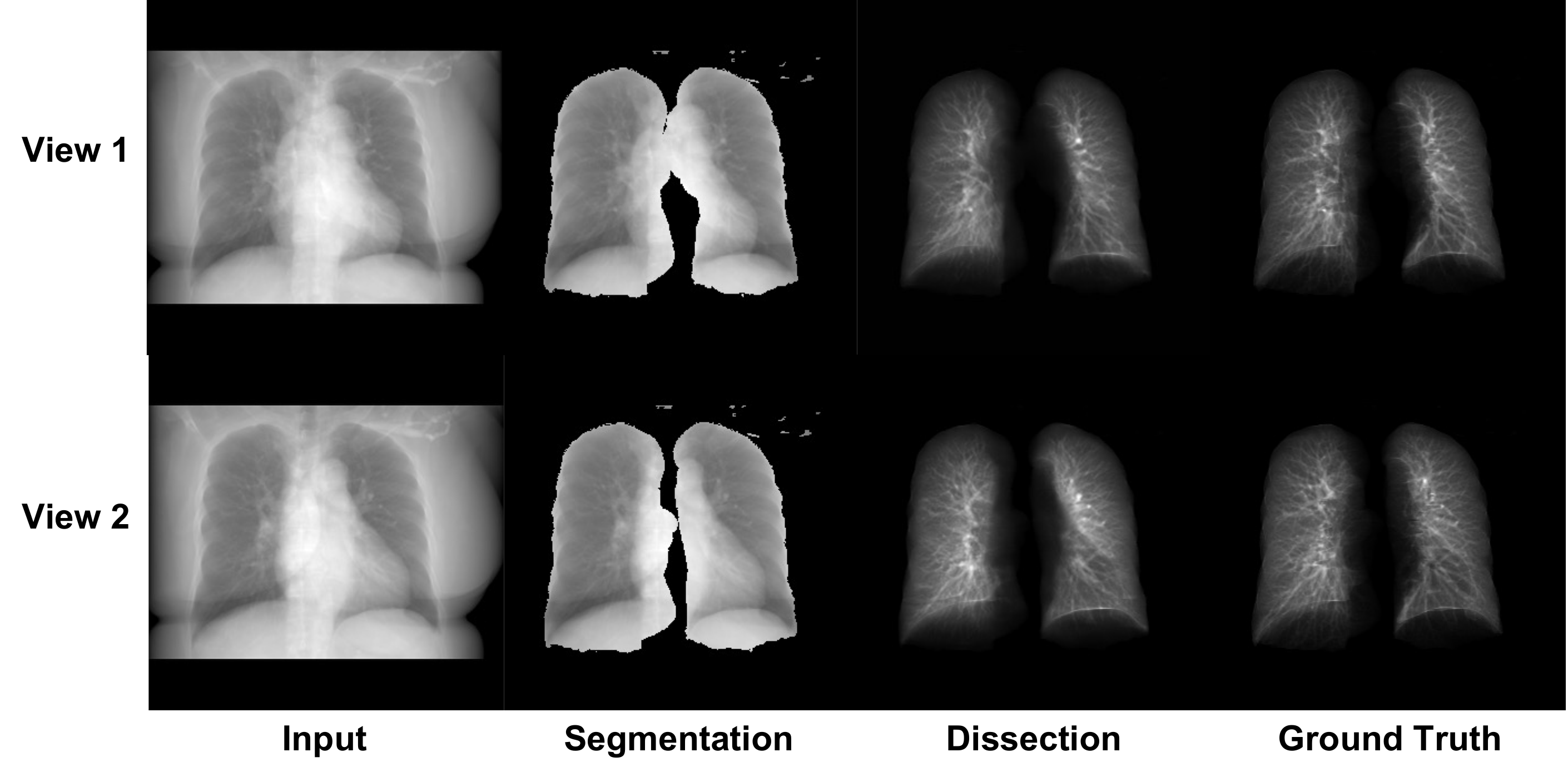}
    \caption{XST testing results from two stereo views of a different patient.}
    \label{fig_result_pat_xst}
\end{figure}

\begin{figure}
    \centering
    \includegraphics[width=1.0\textwidth]{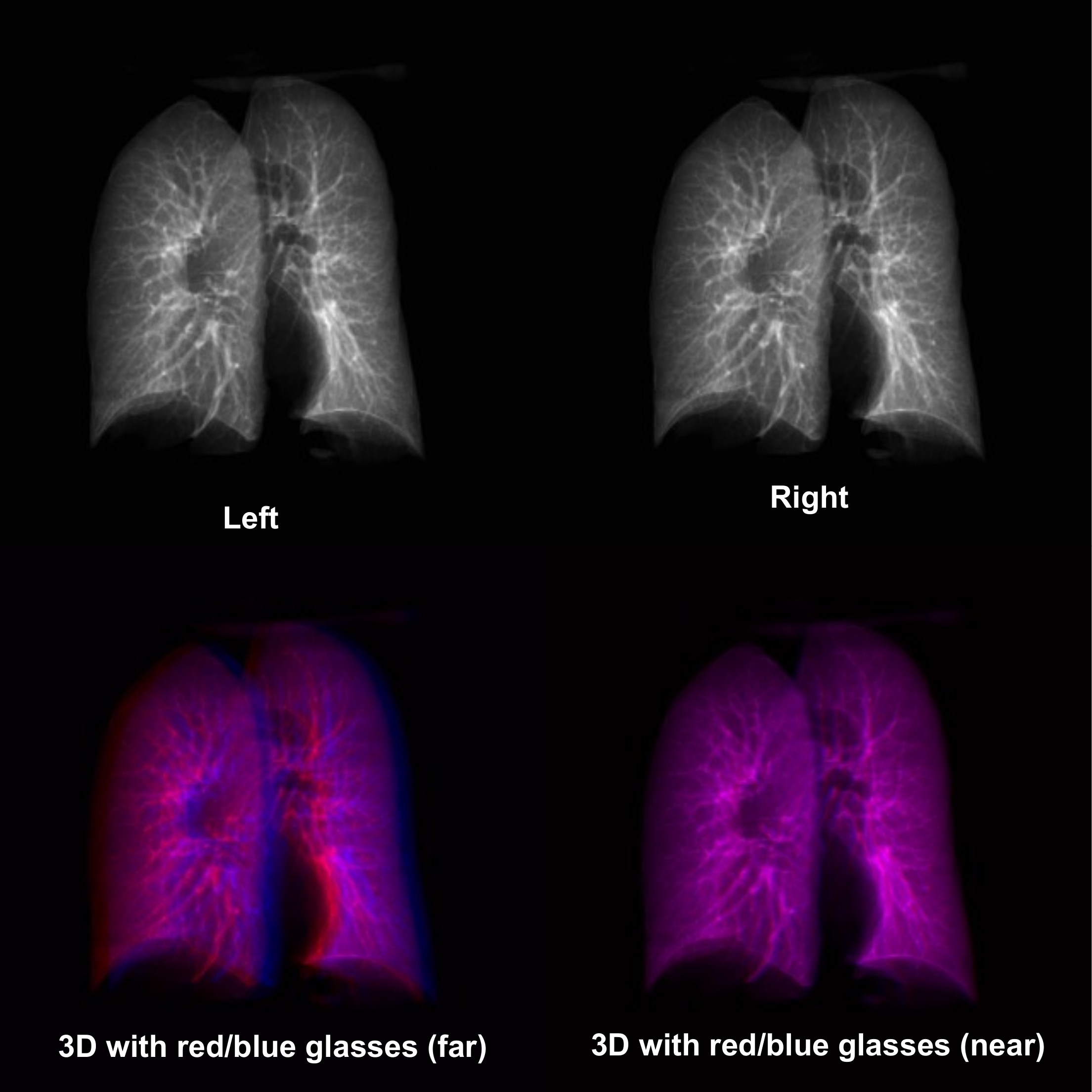}
    \caption{XST image pair and rendering for stereo-viewing.}
    \label{fig_result_stereo}
\end{figure}

Then, we evaluated the feasibility of the proposed XST-Net for X-ray stereoscopic imaging on the CT-lung dataset.
We first evaluated the joint dissection results from two stereo radiograms collected at two new angles of the same patient and then generalized the stereo-imaging technology to different patients.
Our representative results are shown in Fig. \ref{fig_result_angle_xst} and \ref{fig_result_pat_xdt} respectively.
Comments can be made similar to those given in Subsection \ref{sec_result_xdt}. Briely speaking, XST-Net achieved very promising results for predicting different angles, and although results extended to different patients were not as excellent as the counterparts from the same patient, they are still impressive. We believe that the main reason is the limited size of our used dataset.
In addition, we have found that the dissection networks are quite robust to segmentation results, geometric parameters, and image noise.

Finally, we generated 3D perception by overlapping the left-eye and right-eye images in red and blue channels and then viewing both through a pair of red/cyan glasses. Fig. \ref{fig_result_stereo} shows stereo images and two 3D images adjusted with different geometric parameters, as discussed in Subsection \ref{sec_system}. Readers can enjoy watching the 3D lungs through red/cyan glasses (you may need some visual adaptation to see the 3D scene).

\section{Discussions and Conclusion}

As reviewed in the Introduction, while deep reconstruction from few projections was reported before by several groups, no efforts were made in their studies to extract target organs and tissues. On the other hand, there were reports on digital removal of target structures (mostly, ribs) from radiographs, they paid attention to neither stereo viewing or tomographic visualization. Our approach is innovative in the synergistic integration of dissection, stereo viewing, and tomographic reconstruction in the deep learning framework. Furthermore, our approach can be either individualized or population-based to improve radiomics potentially and contribute to precision medicine significantly from stereo-angles.

As mentioned earlier, radiography and CT are two most popular forms of x-ray imaging, and tomosynthesis is the midway between these extreme. From radiography, through tomosynthesis, limited angle tomography and few-view tomography, to CT, overlapped structures are gradually unraveled to produce increasingly more clarify for medical imaging. In this context, our proposed XDT/XST approach is an intelligent way to remove overlapping structures using deep networks, instead of radiation dose. Such digital dissection has the same effect of extracting signals from intervening background components, holding a great potential for real-world x-ray imaging. Indeed, over the past two hundred years, x-ray radiography is exclusively performed in the 2D display without leveraging human stereo perception. With the rapid development of virtual reality techniques, such as ''metaverse'', stereoscopic devices will be much improved and popularized in the near future. Hence, our XDT/XST should have an impact on radiology practice in the future.

In our initial experiments, we have evaluated the proposed deep learning framework for x-ray dissectography and x-ray stereotography, and suggested the top-level designs of the corresponding X-ray imaging systems.
In this study, we only reported the visual results to establish the feasibility.
To demonstrate the utility of this new approach, XDT and XST networks and systems should be evaluated in specific applications.
For example, digital dissection of an organ or tissue type in either 2D projection or 3D image spaces may improve radiomics analysis.
For instance, the lung dissection results may improve the detection and classification performance of lung diseases \cite{mimic} such as the screening and monitoring of COVID-19 \cite{covid}.
On the other hand, by mixing the machine intelligence and the human intelligence, the proposed x-ray stereotography may help radiologists deliver better diagnostic reports.

It is underlined that a personalized dissectographic workflow may be developed for individualized precision medical imaging. Specifically, assuming a 3D CT volume of a patient is available, we can only use this patient CT volume for data augmentation (various realistic disturbances, e.g., imaging under different conditions (distances and orientations) and manipulating various structures and lesions), imaging simulation and network training. Such a trained model may give the best follow-up performance. 
In principle, personalized dissectography should produce very accurately and reliably separated target organs/tissues than population-based dissectography. In other words, follow-up CT screening of lung cancer could be replaced by our proposed XDT/XST.
At the same time, we believe that population-based dissectography should be also workable, as shown in this study, but a much larger dataset will be needed.

Although we have only focused on separation of the lungs in this study, the proposed x-ray dissectography can be extended for other target organs/tissues, and other imaging modalities such as dark-field and phase-contrast radiography, scintigraphy (also known as a gamma/SPECT scan), and almost all other forms of tomography where separation of targets from background is beneficial for signal-to-noise improvement.
More accurate phantoms, more advanced simulation tools, and more powerful network models can be used for superior digital dissection results.

In conclusion, we have proposed the x-ray dissectography (XDT) and x-ray stereography (XST) systems and methods for improving the utilities of conventional X-ray radiography and digital tomosynthesis. The proposed XDT and XST can dissect a target organ/tissue type from X-ray radiograms with deep learning.
The experimental results clearly demonstrate the feasibility and potential utility of the proposed imaging technology.
In the future, we will continue improving the network model and producing clinically relevant results systematically. Hopefully, the proposed XDT and XST techniques empowered by artificial intelligence may open a new door for traditional X-ray radiography to have new impacts on healthcare.


\bibliographystyle{unsrt}  
\bibliography{references}

\begin{thebibliography}{10}

\bibitem{stes}
Laurens Hogeweg, Clara~I. Sánchez, and Bram van Ginneken.
\newblock Suppression of translucent elongated structures: Applications in
  chest radiography.
\newblock {\em IEEE Transactions on Medical Imaging}, 32(11):2099--2113, 2013.

\bibitem{ribcenterline}
Dijia Wu, David Liu, Zoltan Puskas, Chao Lu, Andreas Wimmer, Christian Tietjen,
  Grzegorz Soza, and S.~Kevin Zhou.
\newblock A learning based deformable template matching method for automatic
  rib centerline extraction and labeling in ct images.
\newblock In {\em 2012 IEEE Conference on Computer Vision and Pattern
  Recognition}, pages 980--987, 2012.

\bibitem{li2019encoding}
Zeju Li, Han Li, Hu~Han, Gonglei Shi, Jiannan Wang, and S.~Kevin Zhou.
\newblock Encoding ct anatomy knowledge for unpaired chest x-ray image
  decomposition, 2019.

\bibitem{peng2020xraysyn}
Cheng Peng, Haofu Liao, Gina Wong, Jiebo Luo, Shaohua~Kevin Zhou, and Rama
  Chellappa.
\newblock Xraysyn: Realistic view synthesis from a single radiograph through ct
  priors, 2020.

\bibitem{hbs}
Han Li, Hu~Han, Zeju Li, Lei Wang, Zhe Wu, Jingjing Lu, and S.~Kevin Zhou.
\newblock High-resolution chest x-ray bone suppression using unpaired ct
  structural priors.
\newblock {\em IEEE Transactions on Medical Imaging}, 39(10):3053--3063, 2020.

\bibitem{han2021ganbased}
Luyi Han, Yuanyuan Lyu, Cheng Peng, and S.~Kevin Zhou.
\newblock Gan-based disentanglement learning for chest x-ray rib suppression,
  2021.

\bibitem{enhance}
Ophir Gozes and Hayit Greenspan.
\newblock Lung structures enhancement in chest radiographs via {CT} based
  {FCNN} training.
\newblock {\em CoRR}, abs/1810.05989, 2018.

\bibitem{Ying_2019_CVPR}
Xingde Ying, Heng Guo, Kai Ma, Jian Wu, Zhengxin Weng, and Yefeng Zheng.
\newblock X2ct-gan: Reconstructing ct from biplanar x-rays with generative
  adversarial networks.
\newblock In {\em The IEEE Conference on Computer Vision and Pattern
  Recognition (CVPR)}, June 2019.

\bibitem{shen2019patient}
Liyue Shen, Wei Zhao, and Lei Xing.
\newblock Patient-specific reconstruction of volumetric computed tomography
  images from a single projection view via deep learning.
\newblock {\em Nature biomedical engineering}, 3(11):880--888, 2019.

\bibitem{shen2021geometryinformed}
Liyue Shen, Wei Zhao, Dante Capaldi, John Pauly, and Lei Xing.
\newblock A geometry-informed deep learning framework for ultra-sparse 3d
  tomographic image reconstruction, 2021.

\bibitem{robotic}
Andreas Fieselmann, Jan Steinbrener, Anna~K. Jerebko, Johannes~M Voigt,
  Rosemarie Scholz, Ludwig Ritschl, and Thomas Mertelmeier.
\newblock {Twin robotic x-ray system for 2D radiographic and 3D cone-beam CT
  imaging}.
\newblock In Despina Kontos and Thomas~G. Flohr, editors, {\em Medical Imaging
  2016: Physics of Medical Imaging}, volume 9783, pages 128 -- 133.
  International Society for Optics and Photonics, SPIE, 2016.

\bibitem{dukesim}
Ehsan Abadi, Brian~P. Harrawood, Shobhit Sharma, Anuj~J. Kapadia, William~Paul
  Segars, and Ehsan Samei.
\newblock Dukesim: A realistic, rapid, and scanner-specific simulation
  framework in computed tomography.
\newblock {\em IEEE Transactions on Medical Imaging}, 38:1457--1465, 2019.

\bibitem{catsim}
Bruno~De Man, Samit Basu, Naveen Chandra, Bruce Dunham, Peter Edic, Maria
  Iatrou, Scott McOlash, Paavana Sainath, Charlie Shaughnessy, Brendon Tower,
  and Eugene Williams.
\newblock {CatSim: a new computer assisted tomography simulation environment}.
\newblock In Jiang Hsieh and Michael~J. Flynn, editors, {\em Medical Imaging
  2007: Physics of Medical Imaging}, volume 6510, pages 856 -- 863.
  International Society for Optics and Photonics, SPIE, 2007.

\bibitem{xcat}
WP~Segars and et~al.
\newblock 4d xcat phantom for multimodality imaging research.
\newblock {\em Medical Physics}, 2010.

\bibitem{mosdata}
S.~P. Morozov, A.~E. Andreychenko, N.~A. Pavlov, A.~V. Vladzymyrskyy, N.~V.
  Ledikhova, V.~A. Gombolevsky, Ivan~A. Blokhin, P.~B. Gelezhe, A.~V. Gonchar,
  and Valeria Chernina.
\newblock Mosmeddata: Chest {CT} scans with {COVID-19} related findings
  dataset.
\newblock {\em CoRR}, abs/2005.06465, 2020.

\bibitem{orgdata}
Patrick Bilic and et~al.
\newblock The liver tumor segmentation benchmark (lits).
\newblock {\em CoRR}, abs/1901.04056, 2019.

\bibitem{astra}
Wim van Aarle, Willem~Jan Palenstijn, Jeroen Cant, Eline Janssens, Folkert
  Bleichrodt, Andrei Dabravolski, Jan~De Beenhouwer, K.~Joost Batenburg, and
  Jan Sijbers.
\newblock Fast and flexible x-ray tomography using the astra toolbox.
\newblock {\em Opt. Express}, 24(22):25129--25147, Oct 2016.

\bibitem{niu2021suppression}
Chuang Niu, Fenglei Fan, Weiwen Wu, Mengzhou Li, Qing Lyu, and Ge~Wang.
\newblock Suppression of independent and correlated noise with similarity-based
  unsupervised deep learning, 2021.

\bibitem{mimic}
Alistair E.~W. Johnson and et. al.
\newblock Mimic-cxr, a de-identified publicly available database of chest
  radiographs with free-text reports.
\newblock {\em Scientific Data}, 6, 2019.

\bibitem{covid}
L.A. Rousan and et. al.
\newblock Chest x-ray findings and temporal lung changes in patients with
  covid-19 pneumonia.
\newblock {\em BMC Pulm Med}, 6, 2019.

\end{thebibliography}

\end{document}